\documentclass{aastex6}

\begin{document}

\title{Probing WHIM around Galaxy Clusters with Fast Radio Bursts and
the Sunyaev-Zel'dovich effect}


\author{Yutaka Fujita} \affil{Theoretical Astrophysics,
Department of Earth and Space Science, Graduate School of Science, Osaka
University, 1-1 Machikaneyama-cho, Toyonaka, Osaka 560-0043, Japan}

\author{Takuya Akahori} \affil{Graduate School of Science and
Engineering, Kagoshima University, Kagoshima 890-0065, Japan}

\author{Keiichi Umetsu} \affil{Institute of Astronomy and Astrophysics,
Academia Sinica, P. O. Box 23-141, Taipei 10617, Taiwan}

\author{Craig L. Sarazin} \affil{Department of Astronomy, University of
Virginia, P.O. Box 400325, Charlottesville, VA 22904-4325, USA}

\and

\author{Ka-Wah Wong\altaffilmark{1}} \affil{Eureka Scientific, Inc.,
2452 Delmer Street Suite 100, Oakland, CA 94602-3017, USA}

\altaffiltext{1}{Department of Physics and Astronomy, Minnesota State
University, Mankato, MN 56001, USA}

\begin{abstract}
We propose a new method to probe the Warm Hot Intergalactic Medium
(WHIM) beyond the virial radius ($R_{200}$) of a cluster of galaxies,
where X-ray observations are not easily achievable. In this method, we
use dispersion measures (DMs) of Fast Radio Bursts (FRBs) that appear
behind the cluster and the Sunyaev-Zel'dovich (SZ) effect towards the
cluster. The DMs reflect the density of the intracluster medium (ICM)
including the WHIM. If we observe a sufficient number of FRBs in the
direction of the cluster, we can derive the density profile from the
DMs. Similarly, we can derive the pressure profile from the SZ
effect. By combining the density and the pressure profiles, the
temperature profile can be obtained. Based on mock observations of
nearby clusters, we find that the density of the WHIM can be determined
even at $> 2\: R_{\rm 200}$ from the cluster center when FRB
observations with the Square Kilometre Array (SKA) become available. The
temperature can be derived out to $r\sim 1.5\: R_{\rm 200}$, and the
radius is limited by the current sensitivity of SZ observations.
\end{abstract}




\section{Introduction} 
\label{sec:intro}

It has been predicted that a diffuse warm hot intergalactic medium
(WHIM) at temperatures $T\sim 10^5$--$10^7$~K contains $\sim $50\% of
the baryons in the universe at low redshifts
\citep[e.g.][]{1999ApJ...514....1C,2001ApJ...552..473D}. Although a
number of surveys have been conducted to constrain the WHIM
\citep[e.g.][]{2005Natur.433..495N,2007PASJ...59S.339T,fuj08b,2008ApJ...679..194D,2008ApJS..177...39T},
they have detected only a fraction of the predicted amount of the
WHIM. Some of the WHIM is expected to exist in the outskirts of galaxy
clusters. This WHIM gradually falls into the clusters and is heated at
accretion shocks \citep[e.g.][]{ryu03a}. Thus, we can understand the
process in which the WHIM turns into the hot intracluster medium (ICM)
in the clusters by exploring the WHIM in this region. The outskirts of
clusters have been investigated in X-rays especially with {\it Suzaku}
(e.g.
\citealt{fuj08b,rei09a,2009MNRAS.395..657G,2010ApJ...714..423K,2010PASJ...62..371H,2011PASJ...63S1019A,sim11a,2012MNRAS.422.3503W,2012PASJ...64...95S,2013ApJ...766...90I,2016ApJ...829...49W},
see a recent review by \citealt{rei13a}). Many of these observations
show that the entropy of the ICM in the outskirts of massive clusters is
smaller than that predicted by numerical simulations
(e.g. \citealt{voi05a} but see \citealt{eck13a}), which may indicate
that the heating is less effective than expected
\citep[e.g.][]{2009ApJ...707.1141W,fuj13b}. Unfortunately, the X-ray
observations are limited to $r\lesssim R_{200}$, where $R_{200}$ is the
radius at which the cluster density is 200 times the critical density at
that redshift and is often regarded as the virial radius of the cluster
\citep[e.g.][]{nav96a}. In order to understand the heating process,
observations of the WHIM at $r\gtrsim R_{200}$ are crucial. Although the
signatures of the (thermal) Sunyaev-Zel'dovich (SZ) effect
\citep{1972CoASP...4..173S} have been detected at $r\gtrsim R_{200}$ for
a number of clusters \citep{pla13d,pla15b,pla15a}, it gives us only the
information on the pressure profiles, and the density and the
temperature profiles are not obtained separately.

Fast Radio Bursts (FRBs) are bright, unresolved, millisecond
signals. Although their origin is not clear, their large dispersion
measures (DMs) suggest that they are extragalactic
\citep{2007Sci...318..777L,2012MNRAS.425L..71K,2013Sci...341...53T,2014ApJ...790..101S,2014ApJ...792...19B,2015ApJ...799L...5R,2015MNRAS.447..246P,mas15b}. It
has been proposed that the DMs of transient objects such as gamma-ray
bursts (GRBs) and FRBs can be used to probe the WHIM along the line of
sight \citep{2003ApJ...598L..79I,2004MNRAS.348..999I}. In this study, we
propose a new idea to explore the WHIM by combining the DMs of FRBs and
the SZ effect. We show that this method will reveal the properties of
the WHIM at $r\gtrsim R_{200}$ when observations of FRBs with the Square
Kilometre Array (SKA) become available. We use $H_0=70\:\rm km\:
s^{-1}\: Mpc^{-1}$, $\Omega_m=0.3$, and $\Omega_\Lambda=0.7$.

\section{Dispersion measures of FRBs}
\label{sec:DM}

So far at least 17 FRBs have been identified, and a catalogue compiled
from the published literature is
available\footnote{http://astronomy.swin.edu.au/pulsar/frbcat/}. Their
DMs are $\sim 400$--$1600\rm\: cm^{-3}\: pc$. Although the values of the
DMs may depend on the distance to the FRBs, we treat that as a random
variable for simplicity. From the catalogue, we find that the average of
the DMs is $\rm \langle DM\rangle_{\rm 17}=785\rm\: cm^{-3}\: pc$ and
the square root of the unbiased variance is $\sigma_{\rm DM17}=289\rm\:
cm^{-3}\: pc$.

In this study, we focus on nearby clusters ($z\lesssim 0.1$) and FRBs
that appear behind them. Indeed, the observed large DMs suggests
cosmological origins at high redshifts of $z=0.5$--1
\citep[e.g.][]{aka16b}. If the FRBs have a redshift distribution similar
to that for GRBs, the contribution of foreground FRBs can be ignored
\citep[e.g.][]{2013MNRAS.432.2141C}. Even if the foreground FRBs (and
FRBs residing physically inside the cluster) cannot be ignored, SKA will
be able to identify the host galaxies of the FRBs with its excellent
resolution \citep{mac15a}. Once the hosts are identified, the
determination of their redshifts will be possible. The DM of a single
FRB measured by an observer is
\begin{equation}
\label{eq:DMorg}
 {\rm DM} = \int\frac{n_e}{1+z}dL\:,
\end{equation}
where $n_e$ is the electron density and $z$ is the redshift of the gas
(e.g. equation~(4) in \citealt{den14a}). The integration is
performed along the line of sight. The DM in the direction of a cluster
should consist of
\begin{equation}
\label{eq:DMc}
 \rm DM_c = DM_{FRB} + DM_{IGM} + DM_{ICM} + DM_{MW}\:,
\end{equation} 
where $\rm DM_{FRB}$, $\rm DM_{IGM}$, $\rm DM_{ICM}$, and $\rm DM_{MW}$
are DM contributions from the FRB (and the surrounding medium), the
intergalactic medium (IGM), the ICM, and the Milky Way Galaxy,
respectively. From now on, the term ICM includes the WHIM in the higher
than normal density region around the cluster, unless otherwise
mentioned. Moreover, we represent the average of multiple DMs by
``$\langle\rangle$''. The average $\rm \langle DM\rangle_c$
should be larger than the average when there is no DM contribution from
the ICM:
\begin{equation}
\label{eq:DM}
\rm \langle DM\rangle_{nc} = \langle DM\rangle_{FRB} 
+ \langle DM\rangle_{IGM} 
+ \langle DM\rangle_{MW} \:.
\end{equation}
Note that while the extragalactic contribution, $\rm \langle
DM\rangle_{FRB} + \langle DM\rangle_{IGM}$, may be independent of the
direction to the FRB, $\rm \langle DM\rangle_{MW}$ may be dependent on
direction, because the column density of the interstellar medium (ISM)
of the Milky Way Galaxy varies. In that case, we can use the FRBs that
are detected around the cluster of interest to determine the local value
of $\rm \langle DM\rangle_{MW}$. Above the Galactic plane, we have
estimated that a variation of $\rm DM_{MW}$ in even a huge field of
900~$\rm deg^2$ centered on the Virgo or the Coma cluster is only $\sim
6\rm\: cm^{-3}\: pc$ based on a simple Milky Way model by
\citet{aka13}. Thus, we ignore the Milky Way variation. In the SKA era,
$\rm \langle DM\rangle_{MW}$ would have been well-modeled by observing
numerous pulsars and its dependence on the direction would have been
determined \citep{han14a}, although we cannot know about the electron
density beyond those pulsars and the ISM may be too structured to be
covered by those pulsars especially in the direction of the Galactic
plane. We further assume that the 17 FRBs are not affected by the ICM
since the fraction of the sky covered by rich clusters is small. Thus,
we assume that $\rm \langle DM\rangle_{nc}\sim \rm \langle
DM\rangle_{17}$ and the intrinsic dispersion is $\sigma_{\rm DM}^2=
\sigma_{\rm DM17}^2$.

In order to derive the properties of the ICM, we have to determine $\rm
\langle DM\rangle_{ICM}$. From equation~(\ref{eq:DMorg}), the DM
contribution from the ICM is
\begin{equation}
\label{eq:DMICMorg}
 {\rm DM}_{\rm ICM}
 = \int^{\infty}_{-\infty}\frac{n_{\rm ICM}}{1+z}dL\:,
\end{equation}
where $n_{\rm ICM}$ is the electron density of the cluster. For the
outskirt of a nearby cluster, it is expected to be
\begin{equation}
\label{eq:DMICM2}
 {\rm DM}_{\rm ICM} \sim 200 {\rm\:
cm^{-3}\: pc}
\left(\frac{n_e}{1\times 10^{-4}\rm\: cm^{-3}}\right)
\left(\frac{L_c}{\rm 2\: Mpc}\right)\:,
\end{equation}
where $L_c$ is the size of the cluster. This DM is comparable to
$\sigma_{\rm DM}\sim 289\rm\: cm^{-3}\: pc$, which means that the DM
variation of FRBs cannot be ignored when we estimate $\rm
DM_{ICM}$\footnote{On the other hand, observational errors of DMs for
individual FRBs are much smaller than $\sigma_{\rm DM}$ and can be
ignored (see http://astronomy.swin.edu.au/pulsar/frbcat/).}. Since
\begin{equation}
\label{eq:DMICM}
 \rm \langle DM\rangle_{ICM}=\langle DM\rangle_c-\langle DM\rangle_{nc}
\end{equation}
(equations~(\ref{eq:DMc}) and~(\ref{eq:DM})), we need to measure both
$\rm \langle DM\rangle_c$ and $\rm \langle DM\rangle_{nc}$. We can
obtain the former by observing FRBs that appear behind the cluster and
averaging their DMs. The latter can be determined by assuming that $\rm
\langle DM\rangle_{nc}$ equals the average DM of FRBs observed outside
clusters (not behind clusters) or $\rm \langle DM\rangle_{out}$. The
error is given by $\delta{\rm \langle DM\rangle_{out}}=\sigma_{\rm
DM}/\sqrt{N_{\rm out}}$, where $N_{\rm out}$ is the number of FRBs
observed there. If we take a sufficiently wide region outside a cluster,
we expect that $\rm \langle DM\rangle_{out}$ is determined with a
negligible error, because SKA will detect enormous number of FRBs
($N_{\rm out}\gg 1$) and $\delta{\rm \langle DM\rangle_{out} =
\sigma_{\rm DM}}/\sqrt{N_{\rm out}}\rightarrow 0$. Thus, we can
precisely derive $\rm \langle DM\rangle_{nc} (= \langle
DM\rangle_{out})$.

Now we can estimate $\rm \langle DM\rangle_{ICM}$ from $\rm \langle
DM\rangle_c$ and $\rm \langle DM\rangle_{nc}$ using
equation~(\ref{eq:DMICM}).  However, if we attempt to derive $\rm
\langle DM\rangle_{ICM}$ for a narrow region in a cluster projected on
the sky, the number of FRBs observed there, $N_{\rm in}$, can be
relatively small, which may cause an error. The error of $\rm \langle
DM\rangle_{ICM}$ depends on those of $\rm\langle DM\rangle_c$ and
$\rm\langle DM\rangle_{nc}$. While the latter can be ignored because
$\rm \delta\langle DM\rangle_{nc} = \delta\langle DM\rangle_{out}
\approx 0$, the former is written as
\begin{equation}
\label{eq:dDMc}
 \delta\rm \langle
DM\rangle_c^2=\delta\langle DM\rangle_{ICM,int}^2+\delta\langle
DM\rangle_{nc,in}^2\:.
\end{equation}
While $\rm \langle DM\rangle_{ICM}$ is not a directly observable
quantity, its intrinsic uncertainty, $\rm\delta\langle
DM\rangle_{ICM,int}$, can be ignored because $\rm DM_{ICM}$ has nothing
to do with the DM variation of FRBs or $\sigma_{\rm DM}$. The error
$\delta\rm \langle DM\rangle_{nc,in}$ is the uncertainty of $\rm \langle
DM\rangle_{nc}$ that is for the FRBs {\it detected inside the narrow
region}. Since $\delta\rm \langle DM\rangle_{nc,in}$ is given by
$\sigma_{\rm DM}/\sqrt{N_{\rm in}}$, it cannot be ignored if $N_{\rm
in}$ is small. Thus, we obtain $\delta{\rm \langle DM\rangle_c =
\delta\langle DM\rangle_{nc,in} = \sigma_{\rm DM}}/\sqrt{N_{\rm in}}$
from equation~(\ref{eq:dDMc}). In summary, $\rm \langle DM\rangle_{ICM}$
can be derived from equation~(\ref{eq:DMICM}) assuming that $\rm \langle
DM\rangle_{nc} = \langle DM\rangle_{out}$, and the error can be
estimated by
\begin{equation}
\label{eq:dDMICM}
 \delta{\rm \langle DM\rangle_{ICM} = \delta\langle DM\rangle_c =
\sigma_{\rm DM}}/\sqrt{N_{\rm in}}\:.
\end{equation}
In particular, if we consider one specific FRB ($N_{\rm in}=1$), the
error is
\begin{equation}
\label{eq:dDMICM1}
\delta{\rm \langle DM\rangle_{ICM}}
 = \sigma_{\rm DM}\:.
\end{equation}
From now on, we discuss $\rm \langle DM\rangle_{ICM}$ rather than $\rm
\langle DM\rangle_c$, because $\rm \langle DM\rangle_{nc}$ is just a
constant, and we refer to $\rm \langle DM\rangle_{\rm ICM}$ as DM for
simplicity.

\section{Mock Observations}

\subsection{Models}
\label{sec:cluster}

For mock observations, we construct a spherical model cluster based on
observations of the Coma cluster ($z=0.023$), because detailed SZ
observations have been made for the Coma cluster. The radius of the
cluster is $R_{200}=2.62$~Mpc \citep{pla13d}. The thermal SZ effect
from the ICM is represented by the Compton $y$ parameter:
\begin{equation}
\label{eq:y}
 y = \int^{\infty}_{-\infty} 
\sigma_{\rm T}n_{\rm ICM}\frac{k_{\rm B}T_{\rm ICM}}{m_e c^2}dL
= \frac{\sigma_{\rm T}}{m_e c^2}\int^{\infty}_{-\infty}P_{\rm ICM}dL\:,
\end{equation}
where $\sigma_{\rm T}$ is the Thomson cross section, $k_{\rm B}$ is the
Boltzmann constant, $n_{\rm ICM}$ is the ICM electron density, $T_{\rm
ICM}$ is the temperature, $m_e$ is the electron mass, $c$ is the speed
of light, and $P_{\rm ICM}\equiv n_{\rm ICM}k_{\rm B}T_{\rm ICM}$ is the
electron pressure. We ignore the SZ effect other than the thermal SZ
effect produced by the cluster. For the SZ effect, we choose {\it
Planck} for the mock observations. The $y$ parameter has been measured
out to $\sim 1.5\: R_{200}$ with {\it Planck}, and we adopt an ICM
profile that reproduces the {\it Planck} observations
\citep{pla13d}. The density profile is
\begin{equation}
\label{eq:nC}
 n_{\rm ICM}^2(r)
= n_0^2\frac{(r/r_c)^{-\alpha}}{[1+(r/r_c)^2]^{3\beta-\alpha/2}}
\frac{1}{[1+(r/r_s)^3]^{\epsilon/3}}\:,
\end{equation}
where $n_0=2.9\times 10^{-3}\:\rm cm^{-3}$, $r_c=0.4$~Mpc, $r_s=0.7$~Mpc,
$\alpha=0$, $\beta=0.57$, and $\epsilon=1.3$. The temperature profile is
\begin{equation}
\label{eq:TC}
 T_{\rm ICM}(r) = T_0\frac{(r/r_c)^{-a}}{[1+(r/r_t)^b]^{c/b}}\:,
\end{equation}
where $T_0=6.9$~keV, $r_t=0.26$~Mpc, $a=0$, $b=3.4$, and $c=0.6$. The
functional forms of equations~(\ref{eq:nC}) and~(\ref{eq:TC}) were
proposed by \citet{vik06a}. Note that $n_{\rm ICM}$ and $T_{\rm ICM}$
are actually degenerate because {\it Planck} observed the $y$ parameter
that depends on $P_{\rm ICM}$ (equation~(\ref{eq:y})). We assume that
equations~(\ref{eq:nC}) and~(\ref{eq:TC}) can be extrapolated over
$\gtrsim 1.5\: R_{200}$.

We consider mock observations of this model cluster with SKA1-MID, which
is sensitive in the same radio band as the Parkes telescope that has
detected most of the FRBs so far. At present, the estimated event rate
of FRBs based on Parkes detections is $10_{-4}^{+6}\times 10^3\rm\:
events\: sky^{-1}\: day^{-1}$ \citep{mac15a} or $7^{+5}_{-3}\times
10^3\rm\: events\: sky^{-1}\: day^{-1}$ \citep{cha16a}. \citet{mac15a}
predicted that the detection rate would be $\sim 200$ times higher with
SKA1-SUR. Since SKA1-SUR has been deferred, we make an estimation based
on SKA1-MID, which has a sensitivity of a few times better than that of
SKA1-SUR\footnote{http://www.astron.nl/phiscc2014/Documents/Surveys/SKA\_PHISCC\_Braun.pdf}. Assuming
that SKA1-MID's detection rate is $\sim 200$--500 times larger than that
of Parkes, the event rate is estimated to be $\sim 8\times
10^5$--$8\times 10^6\rm\: events\; sky^{-1}\: day^{-1}$, which
means $N_{\rm FRB}\sim 0.8$--$8\rm\: events\; deg^{-2}\: hour^{-1}$.  So
far, most FRBs have been detected with Parkes at 1.2--1.5~GHz. In this
band, the field of view (FOV) of SKA1-MID is $S_{\rm FOV}\sim 0.75\rm\:
deg^2$ at 1.4~GHz \citep{dew16a}. Thus, we expect that the number of
FRBs detected in the SKA's FOV is $N_{\rm FOV}\sim S_{\rm FOV} N_{\rm
FRB} \sim 0.6$--$6\rm\: events\: hour^{-1}$.  The required exposure time
in hours per FOV to detect FRBs at a rate of $n_{\rm SKA} (\rm events\:
deg^{-2})$ is $T_{\rm exp} = n_{\rm SKA}/N_{\rm FRB} = n_{\rm
SKA}/(N_{\rm FOV}/S_{\rm FOV})$.  We can fairly easily achieve a
detection rate of $n_{\rm SKA}=5$--$20\rm\: events\: deg^{-2}$ with a
reasonable exposure time. For example, the required time is $T_{\rm
exp}\sim 0.6$~hour for $n_{\rm SKA}=5\rm\: events\: deg^{-2}$ and
$N_{\rm FOV}= 6\rm\: events\: hour^{-1}$, and $T_{\rm exp}\sim 25$~hours
for $n_{\rm SKA}=20\rm\: events\: deg^{-2}$ and $N_{\rm FOV}= 0.6\rm\:
events\: hour^{-1}$.


Since the apparent size of the Coma cluster may be too large ($\sim
54\rm\; deg^2$ for $<3\: R_{200}$) to cover with SKA1-MID with a
reasonable observation time, we consider a mock cluster located at
$z=0.07$ ($\sim 6.6\rm\; deg^2$ for $<3\: R_{200}$), because massive
clusters like the Coma cluster have been identified around that redshift
\citep[e.g.][]{rei02a}.

\begin{figure}
\begin{center}
\includegraphics[width=100mm]{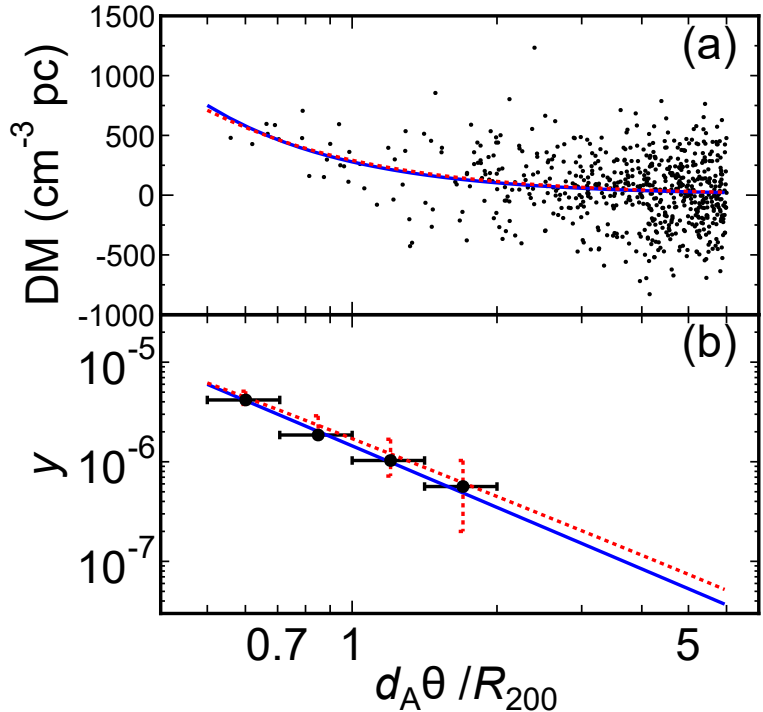} \caption{(a) Dots are a
realization of simulated DMs with statistical errors when $n_{\rm
SKA}=20\rm\: events\: deg^{-2}$, and the solid blue line is the
power-law fit.  The dotted red line is the input-model DM profile, ${\rm
DM}_{\rm ICM}(\theta)$. (b) Dots are a realization of simulated $y$
parameters with statistical errors, and the solid blue line is the
power-law fit. The dotted red line is the input-model $y$ profile,
$y(\theta)$. Vertical bars around the dotted line represent the
$1\:\sigma$ uncertainty $\sigma_{y,i} = \sqrt{C_{ii}}$ at each radial
bin.\label{fig:DMydata}}
\end{center}
\end{figure}

\begin{figure}
\begin{center}
\includegraphics[width=100mm]{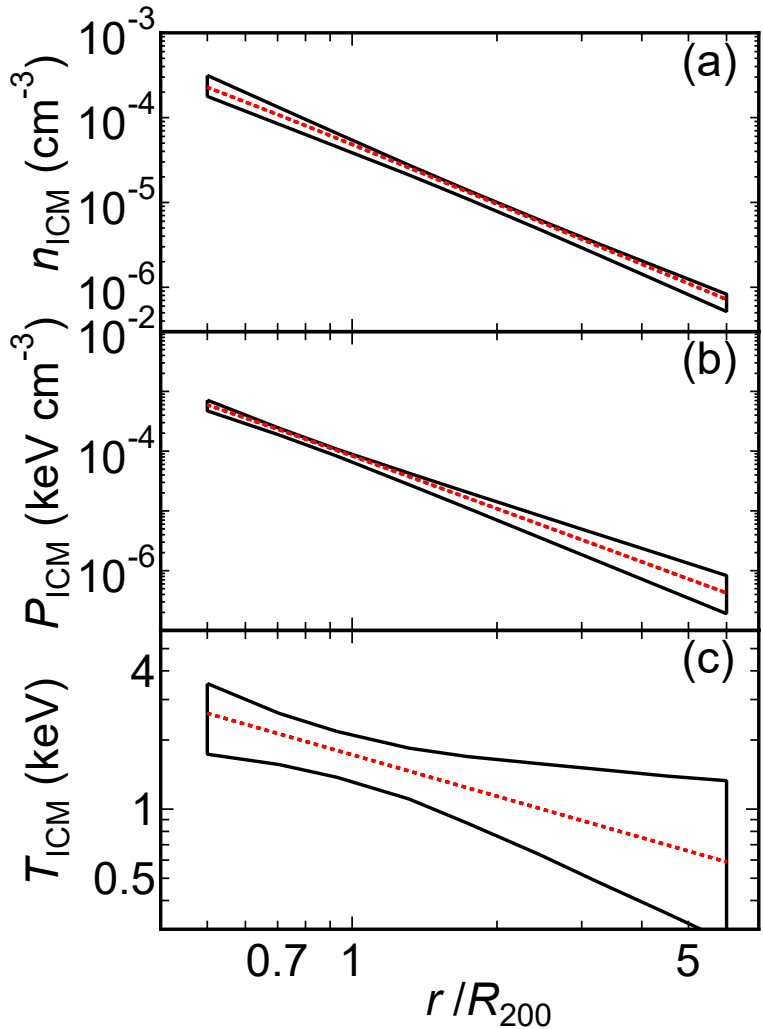} \caption{Solid ribbons are the
reproduced profiles with $1\;\sigma$ errors for (a) ICM density (b)
pressure, and (b) temperature when $n_{\rm SKA}=20\rm\: events\:
deg^{-2}$. Input-model profiles (equations~(\ref{eq:nC})
and~(\ref{eq:TC})) are shown by dotted lines. \label{fig:nPT_fit}}
\end{center}
\end{figure}

\begin{figure}
\begin{center}
\includegraphics[width=100mm]{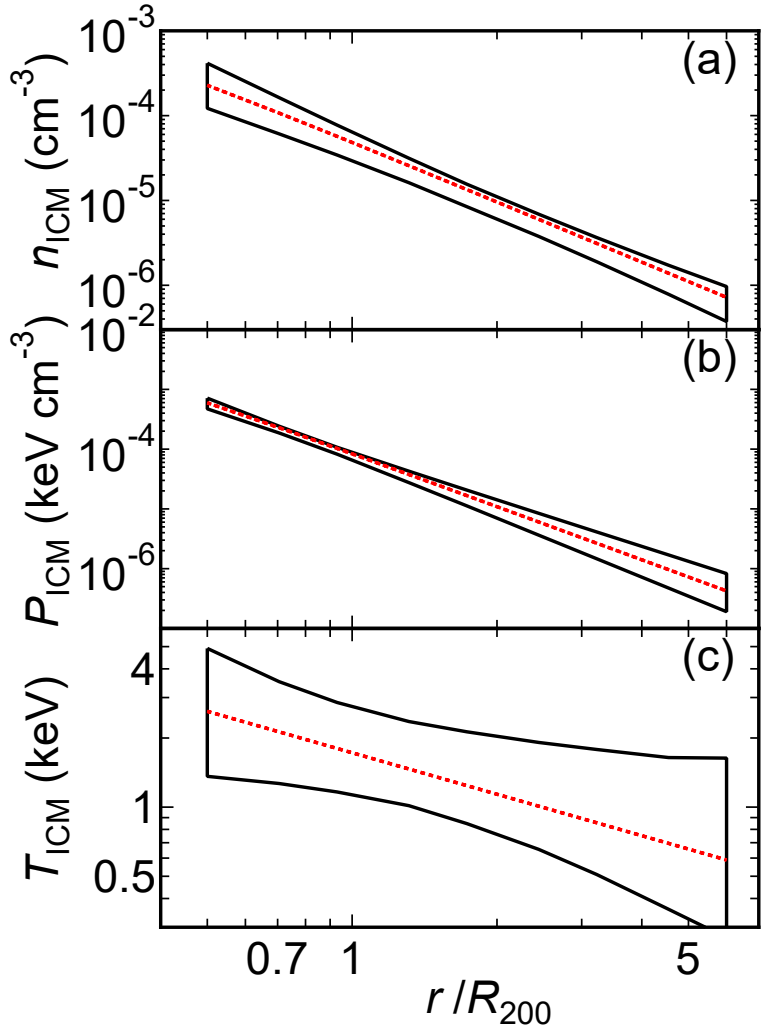} \caption{Same as
Fig.~\ref{fig:nPT_fit} but when $n_{\rm SKA}=5\rm\: events\:
deg^{-2}$. The normalization ${\rm DM_{fit,0}}$ is not fixed in the
fits. \label{fig:nPT_fit_s}}
\end{center}
\end{figure}

\begin{figure}
\begin{center}
\includegraphics[width=100mm]{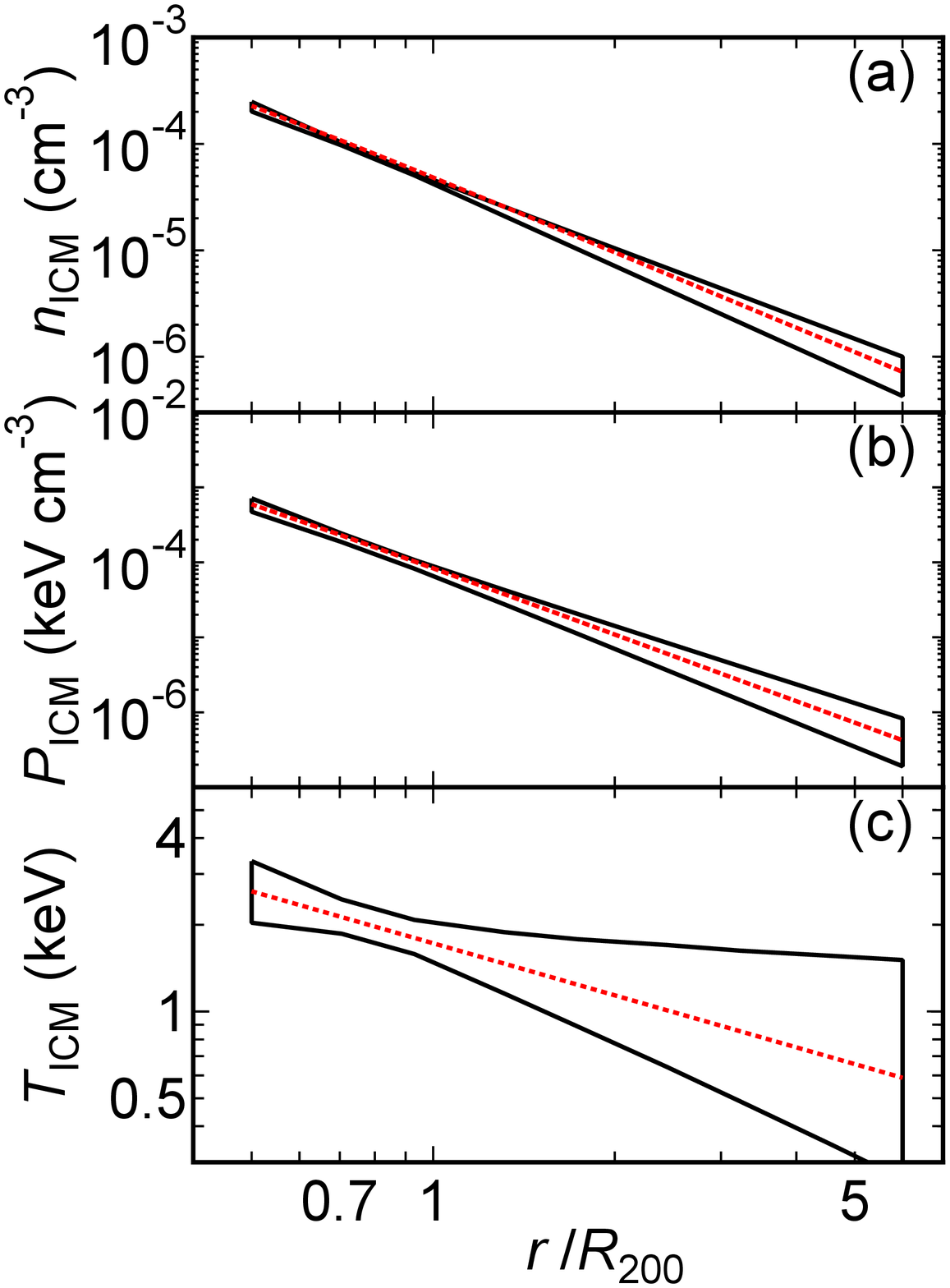} \caption{Same as
Fig.~\ref{fig:nPT_fit_s} but when ${\rm DM_{fit,0}}$ is fixed in the
fits. \label{fig:nPT_fit_q}}
\end{center}
\end{figure}

\subsection{Results of Forward Modeling}
\label{sec:result}

We perform mock observations based on forward modeling, in which
comparison between models and data is made in the data space (DM and
$y$). The forward modeling is appropriate even when the S/N of
observations is not particularly high. In the Appendix, we show the
results of backward modeling (Abel transform), in which comparison is
made in the model space (the ICM density, pressure, and
temperature). Results of the forward and backward modelings should be
consistent within the uncertainty if the S/N is large enough. The
statistical errors are small enough that non-linear effects are not
expected to bias the results. Thus, we ignore the systematic bias
effects.

Using the ICM profile explained in the previous section, we construct
mock DM data and $y$ profiles. For the DM data, we focus on the region
of $r_{\rm min}=\theta_{\rm min} d_{\rm A}=0.5\: R_{200}<\theta d_{\rm
A} <r_{\rm max,DM}=\theta_{\rm max,DM} d_{\rm A}=6\: R_{200}$, where
$\theta$ is the angle from the cluster center, and $d_{\rm A}$ is the
angular diameter distance to the cluster.  For the $y$ parameter, we
consider the region of $r_{\rm min} <\theta d_{\rm A} <r_{\rm
max,y}=\theta_{\rm max,y} d_{\rm A}=2\: R_{200}$, because significant
signals of the SZ effect have been detected at $\lesssim 2\: R_{200}$
\citep{pla13a,pla13d}. Moreover, the influence of {\it Planck's} beam
convolution ($\mathrm{FWHM}=10\arcmin$) can be ignored because $r_{\rm
min}/d_{\rm A}=\theta_\mathrm{min} \sim 30\arcmin$ is much larger than
the FWHM.

We randomly distribute FRBs on the sky with $n_{\rm SKA}=20\rm\:
events\: deg^{-2}$. The input-model DM profile, ${\rm DM}_{\rm
ICM}(\theta)$, can be obtained from
equation~(\ref{eq:DMICMorg}). Following equation~(\ref{eq:dDMICM1}), the
DMs of individual FRBs are randomly perturbed with a Gaussian
distribution, with the dispersion determined by $\sigma_{\rm DM}$. We
note that the actual DMs may not have a Gaussian distribution especially
when the difference of the DMs comes from the redshift of the FRBs. We
obtained $10^3$ different realizations in total. The result of one
realization is shown in Fig.~\ref{fig:DMydata}a. We then fit the DMs
with a power-law:
\begin{equation}
\label{eq:DMfit}
 {\rm DM_{fit}}(\theta)
={\rm DM_{fit,0}}(\theta d_{\rm A}/{r_{\rm min}})^{-s}\:,
\end{equation}
where ${\rm DM_{fit,0}}$ and $s$ are the parameters for the fit.  The
result of a fit and the input-model profile are shown in
Fig.~\ref{fig:DMydata}a. Since the obtained DM profile has a one-to-one
relation to the density profile, it can be converted into the density
profile by the Abel transform (equation~(\ref{eq:nICM})):
\begin{equation}
\label{eq:npow}
 n_{\rm ICM,fit}(r) = \frac{\Gamma((1+s)/2)}{\sqrt{\pi}\:\Gamma(s/2)}
\frac{{\rm DM_{fit,0}}(1+z)}{r_{\rm min}}
\left(\frac{r}{r_{\rm min}}\right)^{-s-1}\:,
\end{equation}
where $\Gamma$ is the gamma function.

The input-model $y$ profile, $y(\theta)$, is derived from
equation~(\ref{eq:y}) and is shown as the dotted line in
Fig.~\ref{fig:DMydata}b. In order to simulate mock {\em Planck}
observations of the model cluster, we logarithmically divide the region
$\theta_\mathrm{min} < \theta < \theta_{\mathrm{max},y}$ into four
concentric annular bins ($N=4$). The observed $y$ value of the $i$-th
radial bin ($i= 1, ..., N$) is expressed as the linear sum of the
bin-averaged $y$ signal, $\langle y_i\rangle$, which can be estimated
from equation~(\ref{eq:y}), and the random noise contribution, $\delta
y_i$.
\begin{equation}
\label{eq:y_i}
 y_i = \langle y_i\rangle + \delta y_i \:.
\end{equation}
Here, $\delta y_i$ ($i= 1, ..., N$) is a noise vector drawn from a
multivariate Gaussian distribution with mean zero and covariance $C_{ij}
\equiv \langle \delta y_i \delta y_j\rangle$ ($i, j = 1, ..., N$). We
express the $i$-th diagonal element of the covariance matrix as
$\sigma_{y,i}=\sqrt{C_{ii}}$, which is the rms noise level at the $i$-th
bin. Noise realizations can be generated as follows: First, we model the
covariance matrix $C_{ij}$ in terms of the noise angular power spectrum
$P_{\ell,\rm noise}$ shown as MILCA-NILC98 F/L data in Fig. 11 of
\citet{pla15a}, where $\ell$ represents the angular multiple. In analogy
with equations~(15) and~(16) of \citet{ume11b}, the covariance matrix is
given by
\begin{equation}
 C_{ij}=\int\frac{\ell d\ell}{2\pi}P_{\ell,{\rm noise}} 
\hat{J}_0(\ell\theta_i)\hat{J}_0(\ell\theta_j)\:,
\end{equation}
where $\theta_i$ is the representative angular radius of the $i$-th bin
given by $y(\theta_i) =\langle y_i\rangle$, and $\hat{J}_0$ is the
Bessel function of the first kind and order zero ($J_0$) that is
averaged over the $i$-th bin (see equation~(16) in \citealt{ume11b}).
Then we perform Cholesky decomposition of the $N\times N$ covariance
matrix as $C = LL^T$, where $L$ is an $N\times N$ lower triangular
matrix. We assign a random noise fluctuation, $\delta y_i$, as in
$\delta y_i=\sum_j L_{ij}\xi_j$, where $\xi_j$ is a random number drawn
from the Gaussian distribution with zero mean and unit variance (see
section 6.1.2 of \citealt{ume10a}). We show the results of one
realization in Fig.~\ref{fig:DMydata}b. We fit $y_i$ ($i=1,..., N$)
with a power-law:
\begin{equation}
\label{eq:yfit}
 y_{\rm fit}(\theta) = y_{\rm fit,0}(\theta d_{\rm A}
/{r_{\rm min}})^{-q}\:,
\end{equation}
where $y_{\rm fit,0}$ and $q$ are the parameters for the fit. The weight
of the fit is $1/\sigma_{y,i}^2$. The result of a fit is shown as the solid
line in Fig.~\ref{fig:DMydata}b. As is the case with DM, the result of
the fit can be converted into the pressure profile
(equation~(\ref{eq:PICM})):
\begin{equation}
\label{eq:Ppow}
 P_{\rm ICM,fit}(r) = \frac{\Gamma((1+q)/2)}{\sqrt{\pi}\:\Gamma(q/2)}
\frac{\sigma_{\rm T}}{m_e c^2}
\frac{{y_{\rm fit,0}}}{r_{\rm min}}
\left(\frac{r}{r_{\rm min}}\right)^{-q-1}\:.
\end{equation}
From equations~(\ref{eq:npow}) and~(\ref{eq:Ppow}), we obtain the
temperature profile $T_{\rm ICM,fit}(r)=P_{\rm ICM,fit}(r)/(k_{\rm B}
n_{\rm ICM,fit}(r))$.  From now on, we omit the suffix `fit' unless
otherwise mentioned.

From $10^3$ Monte--Carlo realizations, we find the reduced $\chi^2$ for
the fit by equation~(\ref{eq:DMfit}) to lie in the range 0.94--1.05
($1\sigma$) with degrees of freedom of $\sim 650$, and that by
equation~(\ref{eq:yfit}) in the range 0.05-0.44 ($1\sigma$) with degrees
of freedom of 2.  Here we note that the small values of the latter are
due to the strong correlation between different radial bins as estimated
from the {\em Planck} $y$ noise angular power spectrum. Note that while
the input-model profiles are not power-laws (equations~(\ref{eq:nC})
and~(\ref{eq:TC})), we fit them power-laws (equations~(\ref{eq:npow})
and~(\ref{eq:Ppow})). However, the power-laws can approximate the
input-model profiles sufficiently well in the cluster outskirt, and
their differences contribute to the reduced $\chi^2$ only by $\sim
3\times 10^{-5}$ for the DM fit and by $\sim 6\times 10^{-4}$ for the
$y$ fit.

At each radius $r$, we derive $1\:\sigma$ ranges of $n_{\rm ICM}$,
$P_{\rm ICM}$, and $T_{\rm ICM}$ and show them in
Fig~\ref{fig:nPT_fit}. The ICM density is determined out to $r\sim 6\:
R_{200}$, and the pressure and the temperature are determined out to
$r\sim 4\: R_{200}$ with uncertainties of a factor of two. However, the
power-law fit of the $y$ data and thus the converted profile of pressure
depend on the $y$ signals available only at $\lesssim 1.5\: R_{200}$
(Fig.~\ref{fig:DMydata}). Thus, the obtained pressure profile for
$\gtrsim 1.5\: R_{200}$ is correct only when the profile can be
represented by a power-law out to the outermost region. On the other
hand, the power-law fit of the DM data depends not only on the inner
region but also on the outer region, because FRBs observed in the outer
region far outnumber those observed in the inner region although their
DMs tend to be small (see Appendix and the similar lengths of the error
bars or $\delta {\rm DM}_i$ in Fig.~\ref{fig:DMy}a). This means that the
ICM density can actually be determined even at $r>2\: R_{200}$.

The detection rate of FRBs and the intrinsic dispersion of their DMs
have large uncertainties at present. Thus, we consider the case where
the detection rate is four times smaller or $n_{\rm SKA}=5\rm\: events\:
deg^{-2}$, which is equivalent to the case where the intrinsic
dispersion is two times larger or $\sigma_{\rm DM}= 2\:\sigma_{\rm
DM17}$. The results are shown in Fig.~\ref{fig:nPT_fit_s}; the errors
become larger by a factor of $\lesssim 2$ than those in
Fig.~\ref{fig:nPT_fit}. One reason for the larger error is the small
number of FRBs detected in the inner region, which leads to the large
uncertainty of $n_{\rm ICM}$ at $r\sim r_{\rm min}=0.5\: R_{200}$
(Fig.~\ref{fig:nPT_fit_s}a). However, the ICM density may be estimated
well at that radius with high spatial resolution SZ measurements such as
MUSTANG-2\footnote{http://www.gb.nrao.edu/mustang/} and
NIKA2\footnote{http://ipag.osug.fr/nika2/Welcome.html}. Thus, we study
the case where we fix the normalization ${\rm DM_{fit,0}}={\rm
DM_{fit}}(\theta_{\rm min})$ in equation~(\ref{eq:DMfit}) at the value
given by the cluster model (section~\ref{sec:cluster}) and we vary only
the index $s$. The results are shown in Fig.~\ref{fig:nPT_fit_q}; the
errors are much reduced compared with Fig.~\ref{fig:nPT_fit_s}
especially at $r\sim R_{200}$. Note that $n_{\rm ICM}$ at $r_{\rm
min}=0.5\: R_{200}$ slightly changes because the normalization of
$n_{\rm ICM}$ depends on the index $s$ (equation~(\ref{eq:npow})).

FRBs may be affected by pulse broadening through scintillation. The
pulse broadening by the IGM is expected to be negligible compared with
that by the ISM of the Milky Way and the FRB's host galaxy
\citep{god97a,mac13a,mas15b}. If the turbulence that is responsible for
the scintillation is a Kolmogorov type and the cutoff scale is the same,
the strength of the broadening (scattering measure; SM) is proportional
to $n_e^2 L_{\rm obj}$, where $L_{\rm obj}$ is the depth of the
intervening medium (e.g. ISM or ICM) along the line of sight
\citep{god97a,mac13a}. For the ICM, the SM is expected to be large in
the inner region of the cluster. At $r\sim 0.5\: R_{200}=1.31$~Mpc,
which is the innermost radius for our analysis, the density is $n_{\rm
ICM}\sim 2\times 10^{-4}\rm\: cm^{-3}$ (Fig.~\ref{fig:nPT_fit}) and the
depth is $L_{\rm ICM}\sim 1.31$~Mpc. For the ISM in the inner halo of
the Milky Way, the density is $n_{\rm ISM}\sim 0.02\rm\: cm^{-3}$ and
the depth is $L_{\rm ISM}\sim 1$~kpc \citep[e.g.][]{kat15c}. Thus, the
ratio of the SM is
\begin{equation}
 \frac{\rm SM_{ICM}}{\rm SM_{ISM}}
= \frac{n_{\rm ICM}^2 L_{\rm ICM}}{n_{\rm ISM}^2 L_{\rm ISM}}\sim 0.1\:.
\end{equation}
This means that the SM associated with the ICM probably is smaller than
that associated with the ISM. Considering the fact that the latter has
inevitably been affecting the pulses of FRBs, the effects of the former
is probably not serious for the detectability of the FRBs. Moreover,
$\rm SM_{ICM}$ becomes even smaller if the cutoff scale of the
turbulence is larger (equations~(30) and (31) in \citealt{mac13a}),
which is likely given the large size of clusters of galaxies.

\section{Discussion}

We have proposed a new method to observe the ICM including WHIM in the
outskirts of galaxy clusters. In this method, DMs of numerous FRBs and
the SZ effect are combined. Since the DMs and the SZ effect give
information on the ICM density and pressure, respectively, we can
estimate the temperature of the ICM from them. From mock observations of
a massive cluster with SKA1-MID and {\it Planck}, we showed that the ICM
density could be determined even at $> 2\: R_{200}$, while the
temperature profile could be derived out to $\sim 1.5\: R_{200}$, and
this maximum radius is limited by the current sensitivity of the SZ
observations \citep{pla13a,pla13d}. We find that the low-$\ell$ excess
($\ell\lesssim 50$) in the {\em Planck} noise power spectrum
$P_{\ell,\mathrm{noise}}$ due to residual foreground contamination
\citep{pla15a} has a non-negligible impact on the simulated $y$ errors
in the cluster outskirts. Hence, it will be important in cluster
outskirt studies to reduce the level of foreground contamination in $y$
maps at large angular scales.

The combination of our method and X-ray observations can be useful.  For
example, clumpiness of the ICM could be studied.  The ICM density and
temperature have been obtained for the interior ($\lesssim R_{200}$) of
many clusters especially with {\it Suzaku} \citep[e.g.][]{rei13a}. The
X-ray surface brightness of a cluster is represented by
\begin{equation}
 S_X\propto \int^{\infty}_{-\infty} n_{\rm ICM}^2 dL\:,
\end{equation}
if the weak temperature dependence is ignored. The density dependence
($n_{\rm ICM}^2$) is different from that for the DM ($n_{\rm ICM}$) in
equation~(\ref{eq:DMICMorg}). This means that if the ICM is moderately
clumpy on a small scale that cannot be resolved by X-ray telescopes,
$S_X$ will be higher than that for smoothly distributed ICM with the
same mass. This does not happen for the DM, which is not dependent on
the clumpiness for a given column density. Thus, $n_{\rm ICM}$
determined by X-ray observations is systematically higher than that
determined by DM observations. Note that while the $y$ parameter depends
linearly on $n_{\rm ICM}$ as does the DM (equation~(\ref{eq:y})), its
behavior may be different from that of the DM. For example, if the clump
size is relatively large and the ICM is in pressure equilibrium on that
scale, the DM values depend on whether the line-of-sight to a FRB
crosses one or more clumps because the column density fluctuates, while
the $y$ parameter is not much affected by the clumping.

The large-scale structure filaments that connect clusters are also
expected to have the WHIM \citep{1999ApJ...514....1C}, although it is
difficult to detect the SZ effect toward them because they are usually
located at $r\gg R_{\rm 200}$. However, the typical density of a
particular filament can be obtained if a sufficient number of FRBs are
detected toward the filament and their DMs are measured. Here the
location of the filament can be inferred from galaxy distributions.  One
problem is that the density $n_{\rm fil}$ is degenerate with the depth
of the filament along the line of sight $L_{\rm fil}$, because ${\rm
DM_{fil}}\propto n_{\rm fil}L_{\rm fil}$ assuming that $n_{\rm fil}$ is
constant. However, this degeneracy can be resolved by an observation of
line emissions from the filament. For example, the WHIM is expected to
radiate O{\small VII} and O{\small VIII} line emissions \citep{yos03a},
and the line surface brightness varies as $S_{\rm L}\propto n_{\rm
fil}^2 L_{\rm fil}$. Thus, the density can be derived as $n_{\rm
fil}\propto S_{\rm L}/{\rm DM_{fil}}$.  The oxygen lines could be
observed with {\it DIOS} \citep{2015arXiv150308405O} and {\it
Athena}\footnote{http://www.the-athena-x-ray-observatory.eu/} in the
future.

In the future, the efficiency of FRB searches may dramatically improve.
If FRBs are found to be bright at lower frequencies, they can be
detected with SKA1-LOW, which has a huge FOV ($20.77 \rm\: deg^{-2}$ at
110~MHz; \citealt{dew16a}). When SKA2 becomes available, the sensitivity
will increase by a factor of 10 compared with
SKA1\footnote{http://astronomers.skatelescope.org/ska2/}, which means
that the required exposure time will be significantly reduced. Moreover,
if the origin of the dispersion, $\sigma_{\rm DM}^2$, is revealed, its
influence on the error, $\delta{\rm \langle DM\rangle_{ICM}}$
(equation~(\ref{eq:dDMICM})), can be decreased by considering an
appropriate correction based on this knowledge.  For example, if the DMs
of FRBs vary with the distance and the positions of the hosts are known
well enough to get a redshift, the distance dependence can be
removed. This could greatly reduce the uncertainty of the DM values, and
make the errors in our method much smaller.  Although we have focused on
individual cluster measurements, this method can be readily generalized
to study a stacked ensemble of high-redshift clusters or group-sized
systems. The Canadian Hydrogen Intensity Mapping
Experiment\footnote{http://chime.phas.ubc.ca/} and The Next Generation
Very Large Array\footnote{https://science.nrao.edu/futures/ngvla} may
also be useful for this study.

\acknowledgments

We thank Ue-Li Pen for useful discussion. This work was supported by
KAKENHI No.~15K05080 (YF), 15H03639 (TA), and~15K17614 (TA). KU
acknowledges support from the Ministry of Science and Technology of
Taiwan through grants MOST 103-2112-M-001-030-MY3 and MOST
103-2112-M-001-003-MY3. CLS was supported in part by Chandra grants
GO5-16131X and GO5-16146X and NASA XMM-Newton grants NNX15AG26G and
NNX16AH23G. KWW was partially supported by Chandra grants GO5-16125X,
GO6-17108X, and NASA ADAP grant NNH16CP10C.

\appendix

\section{Backward Modeling}

In contrast with forward modeling (section~\ref{sec:result}), backward
modeling does not require an ICM model for fits. However, it has a
drawback that we need to use a derivative of the data (see
equations~(\ref{eq:nICM}) and~(\ref{eq:PICM})), which can magnify noise
in the data. Moreover, the errors of the ICM data in the inner region
are affected by those in the outer region (see
equations~(\ref{eq:nICM_num}) and~(\ref{eq:PICM_num})), which may
complicate the comparison between the data and a model. However, it is
instructive to compare the results of the backward modeling with those
of the forward modeling when the ICM profile is rather smooth and the
data have less noise.

\subsection{Abel transform for radial ICM profiles}
\label{sec:Abel}

We assume that clusters are spherically symmetric for the sake of
simplicity. Observable quantities for clusters are often written in the
form of an integration along the line of sight ($L$):
\begin{equation}
\label{eq:abel1}
 f(\theta) = \int^{\infty}_{-\infty} g(L) dL 
= 2\int_{d_{\rm A}\theta}^{\infty} g(r)
\frac{rdr}{\sqrt{r^2 - d_{\rm A}^2 \theta^2}}\:,
\end{equation}
where $\theta$ is the angle from the cluster center projected on the
sky, $r$ is the three-dimensional distance from the cluster center, and
$d_{\rm A}$ is the angular diameter distance to the cluster. Using the
Abel transform, equation~(\ref{eq:abel1}) is converted to be
\begin{equation}
\label{eq:Abel}
 g(r) = -\frac{1}{\pi d_{\rm A}}\int_{r/d_{\rm A}}^{\infty}
\frac{df(\theta)}{d\theta}
\frac{d\theta}{\sqrt{\theta^2-r^2/d_{\rm A}^2}}
\end{equation}
\citep[e.g.][]{yos99a}. In the case of the DM associated with a cluster,
we take $f(\theta)={\rm DM}_{\rm ICM}(\theta)$ and $g(r)=n_{\rm
ICM}(r)/(1+z)$ (equation~(\ref{eq:DMICMorg})). Therefore,
\begin{equation}
\label{eq:nICM}
  n_{\rm ICM}(r) = -\frac{1+z}{\pi d_{\rm A}}\int_{r/d_{\rm A}}^{\infty}
\left[\frac{d}{d\theta}{\rm DM_{\rm ICM}(\theta)}\right]
\frac{d\theta}{\sqrt{\theta^2-r^2/d_{\rm A}^2}}\:.
\end{equation}
The integral in equation~(\ref{eq:nICM}) is numerically evaluated as
\begin{equation}
\label{eq:nICM_num}
 \sum^{N-1}_{i=i_{\rm obs}}
\frac{{\rm DM}_{i+1} 
- {\rm DM}_{i}}
{\sqrt{\theta_{i+1/2}^2-r^2/d_{\rm A}^2}}\:,
\end{equation}
where ${\rm DM}_{i}$ is the average DM of the $i$-th annulus bin, which
corresponds to $\rm \langle DM\rangle_{\rm ICM}$ in
section~\ref{sec:DM}, and $\theta_{i+1/2}\approx
\sqrt{\theta_{i}\theta_{i+1}}$ \citep{yos99a}. The typical angle of the
$i$-the bin, $\theta_i$, is the same as that in
section~\ref{sec:result}. The index $N$ is for the outermost bin and
$i_{\rm obs}$ is for the bin corresponding to $r/d_{\rm A}$.

Equation~(\ref{eq:y}) indicates that by setting $f(\theta)=y(\theta)$
and $g(r)=\sigma_{\rm T}P_{\rm ICM}(r)/(m_e c^2)$ in
equation~(\ref{eq:Abel}), we obtain
\begin{equation}
\label{eq:PICM}
 P_{\rm ICM}(r) = -\frac{m_e c^2}{\pi d_{\rm A}\sigma_{\rm T}}
\int_{r/d_{\rm A}}^{\infty}
\frac{dy(\theta)}{d\theta}
\frac{d\theta}{\sqrt{\theta^2-r^2/d_{\rm A}^2}}\:.
\end{equation}
The integral in equation~(\ref{eq:PICM}) is numerically evaluated as
\begin{equation}
\label{eq:PICM_num}
 \sum^{N-1}_{i=i_{\rm obs}}
\frac{y_{i+1} 
- y_{i}}
{\sqrt{\theta_{i+1/2}^2-r^2/d_{\rm A}^2}}\:,
\end{equation}
where $y_{i}$ is the value of $y$ in the $i$-th bin
(equation~(\ref{eq:y_i})). Combining equations~(\ref{eq:nICM})
and~(\ref{eq:PICM}), we can obtain the temperature profile, $T_{\rm
ICM}(r)=P_{\rm ICM}(r)/(k_{\rm B}n_{\rm ICM}(r))$.

\begin{figure}
\begin{center}
\includegraphics[width=100mm]{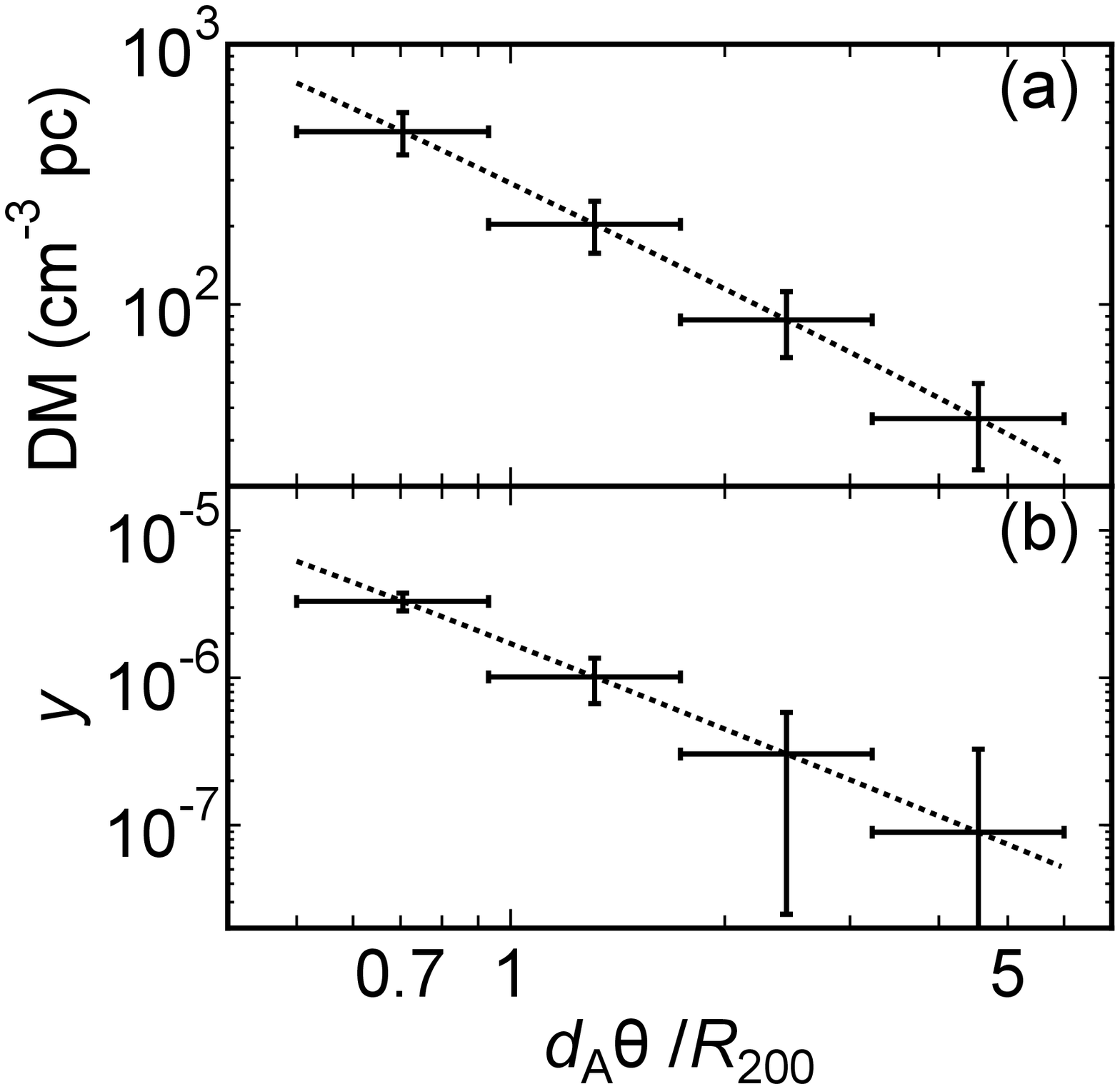} \caption{Original profiles of (a)
the DM and (b) the $y$ parameter when $n_{\rm SKA}=20\rm\: events\:
deg^{-2}$ are shown by the dotted lines. Horizontal error bars show the
size of the bins. Vertical error bars are (a) $\sigma_{\rm
DM}/\sqrt{N_i}$ and (b) $\sigma_{y,i}$. \label{fig:DMy}}
\end{center}
\end{figure}

\begin{figure}
\begin{center}
\includegraphics[width=100mm]{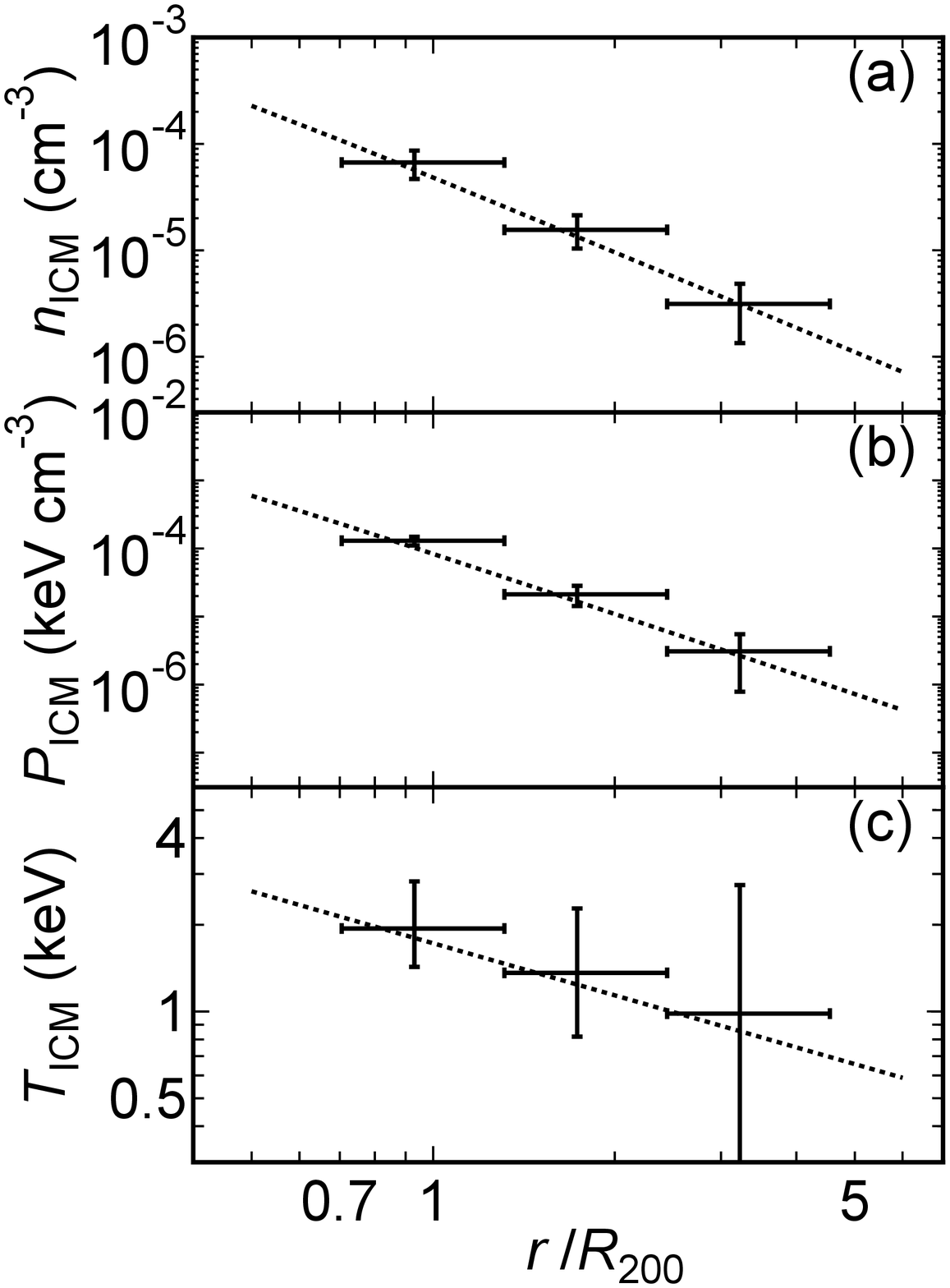} \caption{Crosses are reproduced
profiles of (a) the ICM density and (b) the temperature. Input-model
profiles (equations~(\ref{eq:nC}) and~(\ref{eq:TC})) are shown by dotted
lines. \label{fig:nPT}}
\end{center}
\end{figure}

\subsection{Results}

The model cluster we adopt here is the same as that in
section~\ref{sec:cluster} and we assume $n_{\rm SKA}=20\rm\: events\:
deg^{-2}$. As we did in the forward modeling in
section~\ref{sec:result}, we consider four radial bins in
logarithmically equal intervals ($N=4$). We fix the outer boundary at
$r_{\rm max}=d_{\rm A}\theta_{\rm max}=6\: R_{200}$ and the inner
boundary at $r_{\rm min}=d_{\rm A}\theta_{\rm min}=0.5\: R_{200}$. The
bins are common for DM and $y$. We calculate the assumed model profiles
of ${\rm DM}_{\rm ICM}(\theta)$ and $y(\theta)$ using
equations~(\ref{eq:DMICMorg}) and~(\ref{eq:y}) and the results are shown
by the dotted lines in Fig.~\ref{fig:DMy}. We investigate whether the
input density and temperature profiles are reproduced by the Abel
transform~(equations~(\ref{eq:nICM}) and~(\ref{eq:PICM})). The results
of the Abel transforms are shown in Fig.~\ref{fig:nPT}. The temperature
of the ICM is given by $T_{\rm ICM} = P_{\rm ICM}/(n_{\rm ICM}k_{\rm
B})$. Uncertainties in the results were estimated using Monte Carlo
simulations. At each annulus, the DM is randomly perturbed with a
Gaussian distribution, with an amplitude determined by $\sigma_{\rm
DM}/\sqrt{N_i}$ (equation~(\ref{eq:dDMICM})), where $N_i=n_{\rm SKA}
S_i$ and $S_i$ ($\rm deg^2$) is the solid angle of the $i$-th bin. For
the $y$ parameter, we calculate simulated data with statistical errors,
$y_i$, as we did in section~\ref{sec:result}. We obtained $10^3$
different realizations. Fig.~\ref{fig:nPT} shows that the input-model
profiles are well reproduced by the mock observations. The ICM density
is determined out to $r\sim 3\: R_{200}$ with uncertainties of a factor
of two, while the pressure and the temperature are determined out to
$r\sim 1.5\: R_{200}$. The large errors for the $y$ parameter at $d_{\rm
A}\theta>2\: R_{200}$ (Fig.~\ref{fig:DMy}) limit the maximum radius for
the determination of the pressure and temperature profiles. In
Fig.~\ref{fig:nPT}, the centers of the crosses (reproduced quantities)
do not necessary lie on the the dotted lines (input-model profiles),
because we adopt relatively wide bins. Note that the number of the FRBs
in the outermost bin in Fig.~\ref{fig:DMy} ($d_{\rm A}\theta\sim
3$--$6\: R_{200}$) is $\sim 480$. This means that the number of FRBs
needed to derive the background dispersion measure, $\rm \langle
DM\rangle_{out}$, must be $N_{\rm out}\gg 480$, if one wants to derive
the profiles out to this radius (see section~\ref{sec:DM}).

\end{document}